\documentclass{article}

\usepackage[final]{neurips_2021_ml4ps}

\usepackage[utf8]{inputenc} 
\usepackage[T1]{fontenc}    
\usepackage{hyperref}       
\usepackage{url}            
\usepackage{booktabs}       
\usepackage{amsfonts}       
\usepackage{nicefrac}       
\usepackage{microtype}      
\usepackage{xcolor}         
\usepackage{amsmath}
\usepackage{graphicx}
\usepackage{wrapfig}

\title{Neural Symplectic Integrator with Hamiltonian Inductive Bias for the Gravitational $N$-body Problem}

\author{%
  Maxwell X.~Cai \\
  SURF Corporative / Leiden University \\
  Science Park 140, 1098 XG Amsterdam / Niels Bohrweg 2, 2300 RA Leiden\\
  The Netherlands \\
  \texttt{maxwell.cai@surf.nl}  \\
  \And
  Simon Portegies Zwart \\
  Leiden University \\
  Niels Bohrweg 2, 2300 RA Leiden \\
  \texttt{spz@strw.leidenuniv.nl}
  \And
  Damian Podareanu \\
  SURF Corporative \\
  Science Park 140, 1098 XG Amsterdam \\
  \texttt{damian.podareanu@surf.nl} \\

}

\begin{document}

\maketitle

\begin{abstract}
The gravitational $N$-body problem, which is fundamentally important in astrophysics to predict the motion of $N$ celestial bodies under the mutual gravity of each other, is usually solved numerically because there is no known general analytical solution for $N>2$. Can an $N$-body problem be solved accurately by a neural network (NN)? Can a NN observe long-term conservation of energy and orbital angular momentum? Inspired by Wistom \& Holman's \nocite{1991AJ....102.1528W} symplectic map, we present a neural $N$-body integrator for splitting the Hamiltonian into a two-body part, solvable analytically, and an interaction part that we approximate with a NN. Our neural symplectic $N$-body code  integrates a general three-body system for $10^{5}$ steps without diverting from the ground truth dynamics obtained from a traditional $N$-body integrator. Moreover, it exhibits good inductive bias by successfully predicting the evolution of $N$-body systems that are no part of the training set. 
\end{abstract}

\section{Introduction}
\label{sec:intro}
Deep neural networks (DNNs) have been widely used as universal function approximators to learn a target distribution. Such a property is promising for physics-related applications, whose underlying distributions are typically high-dimensional, computationally expensive, and/or difficult to derive analytically. In astrophysics and particle physics, generative models are often used for Monte-Carlo event generation \citep[e.g.,][]{2018arXiv180203325E,2019arXiv190412846Z}, dimensionality reduction \citep[e.g.,][]{2020AJ....160...45P}, classification \citep[e.g.,][]{2021arXiv210412980F}, and anomaly detection \citep[e.g.,][]{2018MNRAS.476.2117R}.

Despite the success of the aforementioned DNN applications on astrophysical data, the trained models learn to encode representations from the training data, but they do not necessarily understand the underlying physical laws. Making DNNs understand physical laws, therefore, requires special formulations. Physics informed neural networks \citep[PINNs, see][]{2017arXiv171110561R,2017arXiv171110566R,raissi2019physics}, for example, take the form of differential equations and replace the differentiable mathematical function $f$ (which is sometimes unknown or expensive to calculate) by a differentiable neural network $f_{\rm NN}$. In this way, PINNs are trained to approximate the function $f$, and thus differentiating the PINNs will yield the evolution of the underlying system. Moreover, the outputs of $f_{\rm NN}$ becomes more interpretable. This idea is echoed in the recent work by \citet{2019arXiv190601563G}, in which the authors replace the Hamiltonian of a dynamics system with a DNN and successfully apply it to solved problems such as pendulum, mass-spring, and the two-body systems. The resulting architecture, namely Hamiltonian Neural Networks (HNNs), manages to learn and respect the conservation of total energy. Similarly, the Hamiltonian Generative Networks (HGNs) are capable of consistently learning Hamiltonian dynamics from high-dimensional observations (such as images) without restrictive domain assumptions, and the Neural Hamiltonian Flow (NHF), which uses Hamiltonian dynamics to model expressive densities \citep{2019arXiv190913789T}. Various improvements and adoptations have since been made, see e.g., \cite{2019arXiv190913334C, 2019arXiv190912790S, 2020arXiv200304630C,2019arXiv190912077D,2020arXiv201013581F}.


All aforementioned works approximate the full Hamiltonian (or a general function $f$) using a neural network. In this paper, however, we propose a formulation to improve the approximation. We will see that, if the Hamiltonian can be rewritten as an analytically solvable part and a residual part, then approximating only the small residual part with a DNN will improve the prediction accuracy. As an application, we use this idea to construct a neural symplectic integrator for $N$-body simulations \footnote{Source code available at \url{https://github.com/maxwelltsai/neural-symplectic-integrator}.}, based on the formulation by \citet{1991AJ....102.1528W}.  

\section{Methods}
\label{sec:methods}

\subsection{Theoretical Framework}
\label{sec:hamiltonian}
Hamiltonian mechanics is a reformulation of Newtonian mechanics, defined as the sum of potential energy and kinetic energy, i.e., $\mathcal{H} = T + V$, where $T$ is the generalized kinetic energy and $V$ is the generalized potential energy. The function $\mathcal{H}$ is called the Hamiltonian of the system, defined in an abstract phase space $\mathbf{s} = (\mathbf{q}, \mathbf{p})$, where $\mathbf{q} = (q_1, q_2, ..., q_n)$ is the position vector, and $\mathbf{p} = (p_1, p_2, ..., p_n)$ is the momentum vector. The subscript $n$ indicates the dimensionality of the coordinates. The dynamics can be obtained through the following set of differential equations:
\begin{align}
    \mathbf{\dot{q}} \equiv \frac{d\mathbf{q}}{dt} = \frac{\partial \mathcal{H}}{\partial \mathbf{p}} &,& \mathbf{\dot{p}} \equiv \frac{d\mathbf{p}}{dt} = -\frac{\partial \mathcal{H}}{\partial \mathbf{q}}.
\end{align}
Since $\mathcal{H}$ is the total energy, it is a conserved quantity in systems where external perturbations do not exist or can be safely ignored. For a gravitational $N$-body systems, each celestial body experiences the gravity exerted by the other $N-1$ bodies, and the corresponding Hamiltonian is
\begin{equation}
    \mathcal{H} = \sum_{i=0}^{N-1} \frac{p_{i}^2}{2m_i} - G \sum_{i=0}^{N-2} m_i \sum_{j=i+1}^{N-1} \frac{m_j}{r_{ij}}, 
\end{equation}
where $p_i \equiv |\mathbf{p}_i|$ is the magnitude of the momentum vector for particle $i$, and $r_{ij} \equiv |\mathbf{q}_i - \mathbf{q}_j|$ is the distance between particles $i$ and $j$. While there is no known analytic solution for the $N$-body problems with $N>2$, \citet{1991AJ....102.1528W} point out that the Hamiltonian of an $N$-body problem can be naturally split into two parts:
\begin{equation}
    \mathcal{H} = \mathcal{H}_{\rm kepler} + \mathcal{H}_{\rm inter}.
\end{equation}
Here $\mathcal{H}_{\rm kepler}$ represents the Kepler motion of the bodies with respect to the center of mass, and $\mathcal{H}_{\rm inter}$ represents the perturbation of the minor bodies on each other. This splitting is rooted in the perturbation theory, and this formulation allows the integrator to reduce numerical errors by taking the advantage that $\mathcal{H}_{\rm kepler}$ can be solved analytically. Working in Jacobi coordinates, $\mathcal{H}_{\rm kepler}$ can be written as 
\begin{equation}
    \mathcal{H}_{\rm kepler} = \sum_{i=1}^{N-1}\left( \frac{\tilde{p}_{i}^{2}}{2\tilde{m}_i} - G\frac{\tilde{M}_i \tilde{m}_i}{\tilde{r}_i} \right).
\end{equation}
Here $\tilde{m}_i = \left( \sum_{j=0}^{i-1}m_j /  \sum_{j=0}^{i}m_j \right) m_i$ is the normalized mass for body $i$, $\tilde{p}_i = \tilde{m}_i \tilde{v}_i$ is the momentum of body $i$ in Jacobi coordinates, and $\tilde{r}_i$ is the position vector of body $i$ in Jacobi coordinates. The interacting Hamiltonian $\mathcal{H}_{\rm inter}$ takes the form
\begin{equation}
    \mathcal{H}_{\rm inter} = G \sum_{i=1}^{N-1} \frac{m_0 m_i}{\tilde{r}_i} - G \sum_{i=0}^{N-2} m_i \sum_{j=i+1}^{N-1} \frac{m_j}{r_{ij}}.
\end{equation}

The Wisdom-Holman (WH) integrator is a symplectic mapping that employs a kick-drift-kick (or drift-kick-drift) strategy. During the drift step, the integrator assumes that the secondary body orbits the central body (or the center of mass) without experiencing any perturbations. This assumption allows one to propagate the state analytically by solving Kepler's equation. The error for leaving the perturbations from other bodies unaccounted for is subsequently corrected by the kick step, in which the mutual interactions between each body (e.g., planets) are regarded as velocity kicks. A velocity kick $\Delta \mathbf{v} = \mathbf{a}_{\rm p} \Delta t$ accounts for the cumulative effect of all planets due to the perturbing acceleration $\mathbf{a}_{\rm p}$ over a time interval $\Delta t$. Since a kick step only changes the velocity vector without touching the position vector, it ensures that the evolution of position vectors is smooth and differentiable.

\subsection{Neural Interacting Hamiltonian (NIH)}
\label{sec:nn}
$\mathcal{H}_{\rm kepler}$ is an $\mathcal{O}(N)$ operation and $\mathcal{H}_{\rm inter}$ is a more expensive $\mathcal{O}(N^2)$ operation. The former can be solved analytically, but the latter has to be solved numerically. We solve $\mathcal{H}_{\rm kepler}$ analytically but $\mathcal{H}_{\rm inter}$ using a neural network. Following \citet{2019arXiv190601563G}, we use a simple multi-layer perceptron (MLP) backbone network $\mathcal{H}_{\rm inter, \theta}$ to serve as a function approximator of $\mathcal{H}_{\rm inter}$. The network, which we refer to as the neural interacting Hamiltonian (NIH), outputs a single scalar value analog to the total energy of the interaction part. This network is differentiable, and by taking its in-graph gradients with respect to $\mathbf{q}$ and $\mathbf{p}$, we obtain the velocities and the accelerations, respectively. The accelerations are used to calculate the velocity kicks in the WH mapping. Therefore, the networks takes the state vector $(\mathbf{q}, \mathbf{p})$ of the underlying $N$-body system as an input, and generates the time derivative $(\mathbf{\dot{q}}_{\theta}, \mathbf{\dot{p}}_{\theta})$ (every quantity related to the neural network is hereafter denoted with the subscript $\theta$). The loss is calculated as 
\begin{equation}
    \mathcal{L} = = \underbrace{\|\mathbf{\dot{q}} - \mathbf{\dot{q}}_{\theta}\|^2}_{L_2(v)} + \underbrace{\|\mathbf{\dot{p}} - \mathbf{\dot{p}}_{\theta}\|^2}_{L_2(a)} + \underbrace{\| \mathbf{\dot{q}}_{\theta} \times \mathbf{p} - \mathbf{q} \times \mathbf{\dot{p}}_{\theta} \|^2}_{\mathbf{\dot{L}}_{\rm residual}}.
\end{equation}
Here $\mathbf{\dot{q}}$ and $\mathbf{\dot{p}}$ are ground truth velocities and accelerations obtained from a traditional WH integrator. The loss function consists of three parts: the first part $L_2(v)$ aims to minimize the difference between the predicted velocities and the ground truth velocities; the second part $L_2(a)$ aims to minimize the difference between the predicted accelerations and the ground truth acceleration. The third part $\mathbf{\dot{L}}_{\rm residual}$ is essentially the time derivative of the orbital angular momentum $\mathbf{L}$:
\begin{equation}
    \frac{d\mathbf{L}}{dt}  \equiv \frac{d\left( \mathbf{q} \times \mathbf{p} \right)}{dt}  
                   = \frac{d \mathbf{q}}{dt} \times \mathbf{p} + \mathbf{q} \times \frac{d \mathbf{p}}{dt} 
                   = \frac{\partial \mathcal{{H}}_{\theta}}{\partial \mathbf{p}} \times \mathbf{p} - \mathbf{q} \times \frac{\partial \mathcal{H}_{\theta}}{\partial \mathbf{q}}
\label{eq:am_conservation}
\end{equation}
In the absence of external perturbations, there is no external torque and the total angular momentum of an $N$-body system is conserved, hence $d\textbf{L}/dt = 0$. Therefore, minimizing the residual of $d\mathbf{L}/dt$ is equivalent to forcing $\mathcal{H}_{\rm inter, \theta}$ to obey the law of angular momentum conservation. 

In an attempt to improve the backbone network, we also tested a sparsely-connected MLP model \citep[see e.g.,][]{2017arXiv170704780C,2019arXiv190307138P} where $90\%$ of its neurons are randomly zeroed out as a form of regularization. This approach forces a more efficient use of neurons. The sparse MLP model performs slightly better than the vanilla MLP backbone. For simplicity, hereafter we use only the vanilla MLP backbone to carry out all the tests.

\subsection{Simulations, Activation Function, and Training:}
\label{sec:training}
\begin{figure}
    \centering
    \includegraphics[width=0.49\linewidth]{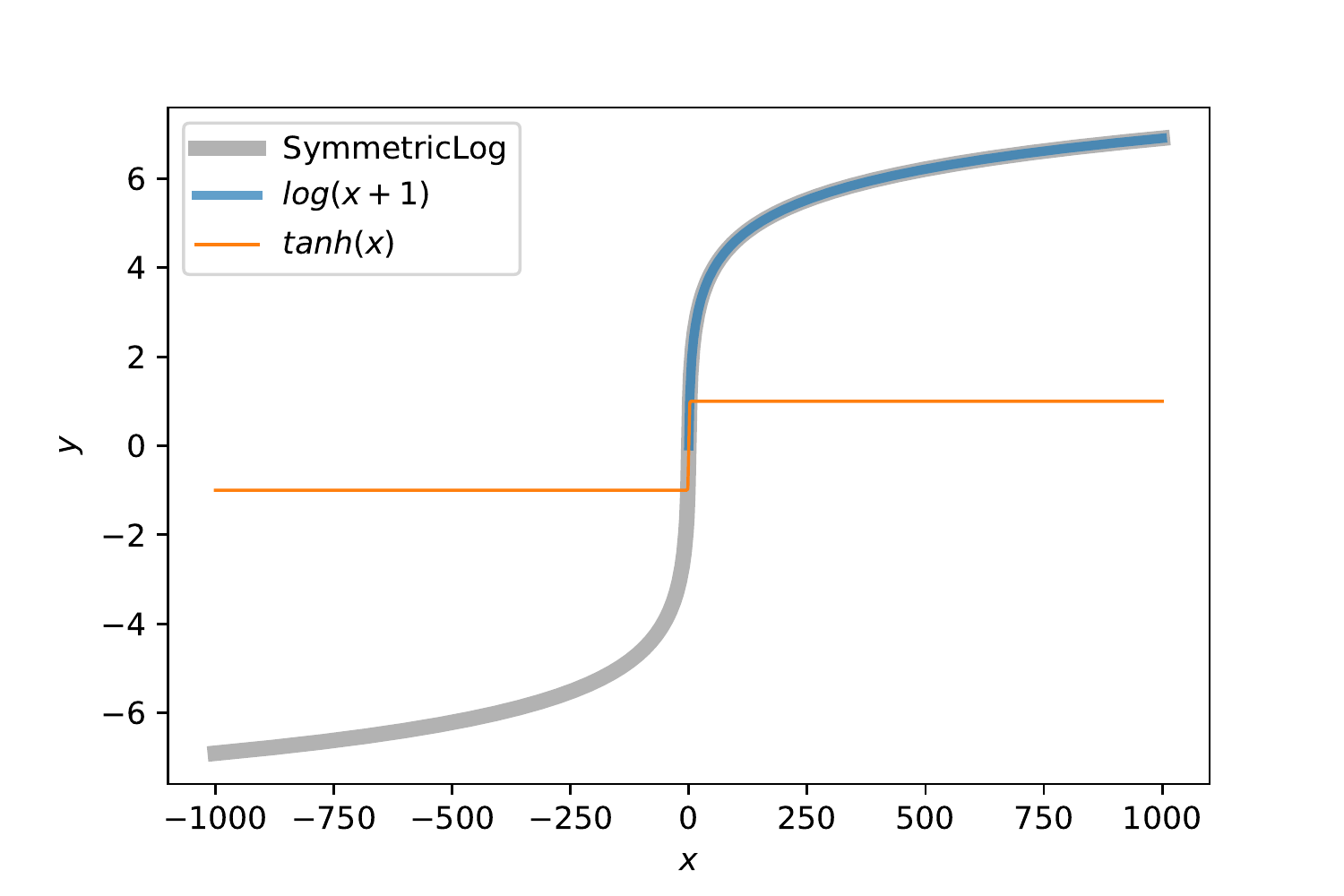} 
    \includegraphics[width=0.49\linewidth]{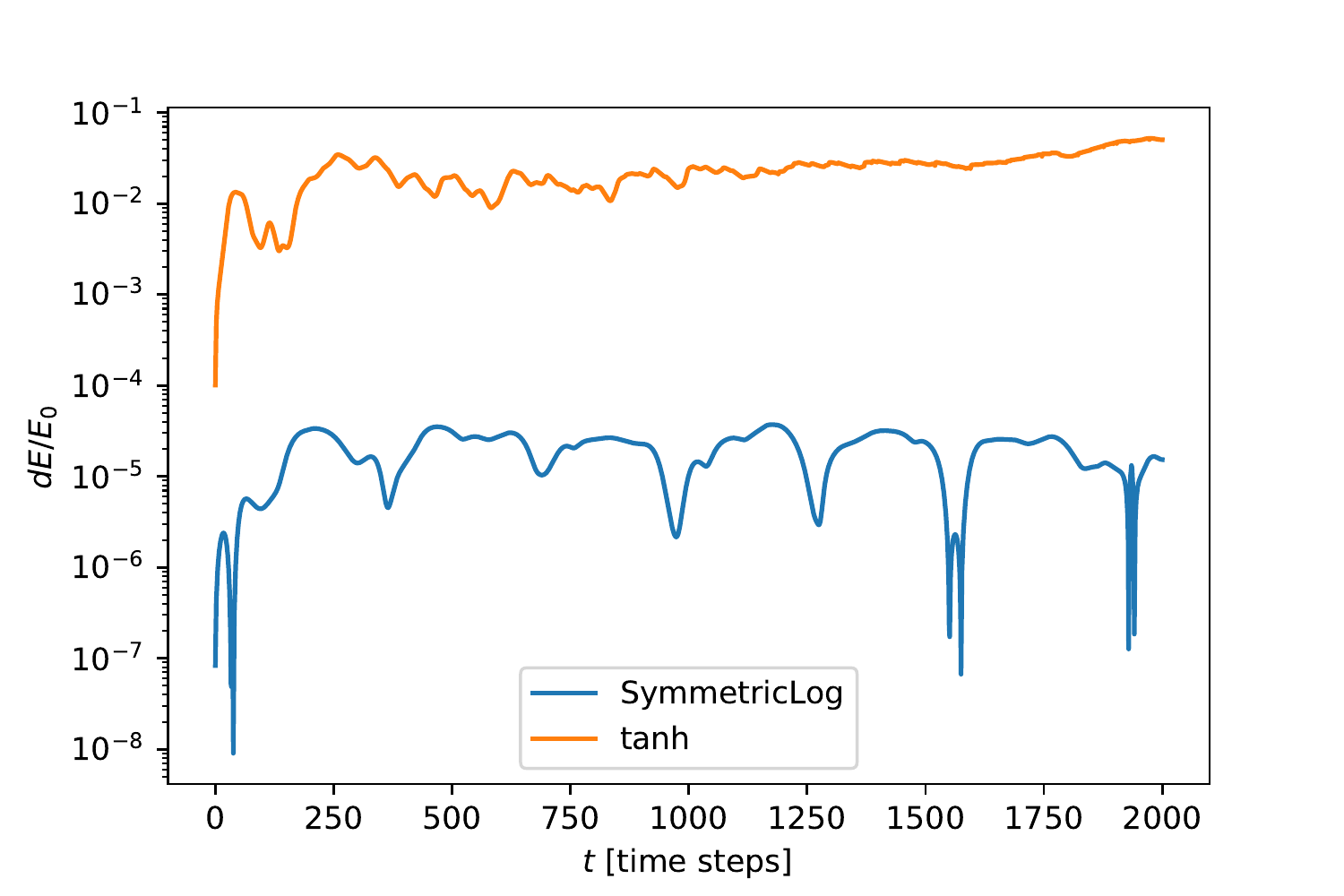}
    \caption{Left: The activation function used in the neural interaction Hamiltonian, as compared to $y = \log(x + 1)$ and $y = \tanh(x)$. Right: Relative integration energy error $dE/E_0$ as a function of time. The \texttt{SymmetricLog} activation performs $\sim 3$ orders of magnitude better in terms of energy conservation compared to \texttt{tanh}.}
    \label{fig:activation}
\end{figure}

{\bf The \texttt{SymmetricLog} Activation Function} Since the neural $N$-body integrator is expected to handle real astrophysical data like any traditional integrators, we need to make sure that $\mathcal{H}_{\rm inter,\theta}$ is capable of handling raw data without the need to perform data pre-processing. Given the vast dynamic ranges in astrophysical data, however, commonly used activation functions, such as \texttt{ReLU}, \texttt{sigmoid}, and \texttt{tanh} have insufficient dynamic range. For example, the \texttt{ReLU} activation function does not activate for negative values, and the \texttt{tanh} activation function saturates to 1 or -1 for absolute  value $>3$. As such, we design a custom activation function \texttt{SymmetricLog}:
\begin{equation}
    f(x) = \tanh(x) \log \left[x \tanh(x) + 1 \right] .
\end{equation}
As shown in Fig.~\ref{fig:activation}, the \texttt{SymmetricLog} activation function behaves similarly as \texttt{tanh} for values close to zero, which delivers strong gradients for backward propagation; for larger values, it behaves like a logarithmic function, which does not saturate or explode the gradients; it is well defined for any real values, making it possible for the neural network to handle raw astrophysical data (unless the quantities are smaller than $10^{-3}$ or larger than $10^4$, in which case data normalization is needed to achieve optimal prediction accuracy). Its symmetry with respect to zero makes it particularly suitable for using in $N$-body simulations because the coordinates of celestial bodies are usually quasi-symmetric with respect to the center of the reference frame. We perform a test of planetary systems with 10 planets orbiting a solar-type star. The semi-major axes of these planets vary from 0.5 au to 10 au, which means that any planets with a semi-major axis greater than $\sim 3$ au will saturate the \texttt{tanh} activation function. \texttt{ReLU} and \texttt{SoftPlus} cannot be used because they cannot generate negative activation outputs. As shown in Fig.~\ref{fig:activation}, \texttt{SymmetricLog} performs $\sim 3$ orders of magnitude better in terms of energy conservation.

{\bf Training} We deliberately train the NIH with a small number of simulations (see Appendix~\ref{appendix:ic} for the initial conditions). In doing so, we wanted to test whether the sub-optimally trained neural network has good inductive bias in transferring the physics laws it learn from the training examples to distinctively different $N$-body systems. We train the neural network with only $N = 3$ systems (there is no NIH for $N = 2$), but the \emph{same} network is subsequently asked to deal with problems with arbitrary $N$. 
For brevity, we refer to a traditional Wisdom-Holman integrator as \texttt{WH}, and the modified Wisdom-Holman integrator with neural interacting Hamiltonian as \texttt{WH-NIH}. We implement \texttt{WH-NIH} using \texttt{PyTorch}, and use the \texttt{Adam} \citep{2014arXiv1412.6980K} optimizer to train the NIH. Given that the training dataset is deliberately made small (roughly 300 MB of \texttt{float32} data), full training can be done on a single GPU or a multi-core CPU within one hour. 


\section{Results}
\label{sec:results}
\subsection{Two-body Problems}
We test the neural integrator with a two-body problem. Since the integrator is trained with three-body simulations, we want to see whether the NIH understands that a two-body problem can be seem as a three-body problem without a perturbing third body. To compare with existing works, we use the same initial condition as \citet{2019arXiv190601563G}. We carry out the simulation for 1,000 time steps, and \texttt{WH-NIH} yields a relative energy error of $dE/E_0 \sim 10^{-9}$ by the end of the simulation. This result is about 7 orders of magnitude more accurate comparing to \citet{2019arXiv190601563G}.

\subsection{The TRAPPIST-1 Exoplanet System}
\textsc{Trappist-1} \citep{2016Natur.533..221G} is an extrasolar planetary system with seven Earth-size planets. Because of its very special orbital architecture (compact super-Earth system with seven known planets orbiting a spectral type M8V star), it has since attracted considerable interests concerning its origin and potential for habitability. We test the \texttt{WH-NIH} integrator using the orbital parameters observed in \textsc{Trappist-1}. The initial conditions of its orbital parameters are obtained from the exoplanets database.\footnote{\url{http://exoplanet.eu}} The innermost planet \textsc{Trappist-1}b has an orbital period of 1.51 days, and its orbital semi-major axis of $a = 0.0011$~au is well below the minimum semi-major axes (0.4 au) in the training dataset, so the \texttt{WH-NIH} integrator has never been trained to handle such an extreme system. For this particular system, we will use an integration time step of $h = 0.365$\,days. 

\begin{figure}
    \centering
    \includegraphics[width=1.0\linewidth]{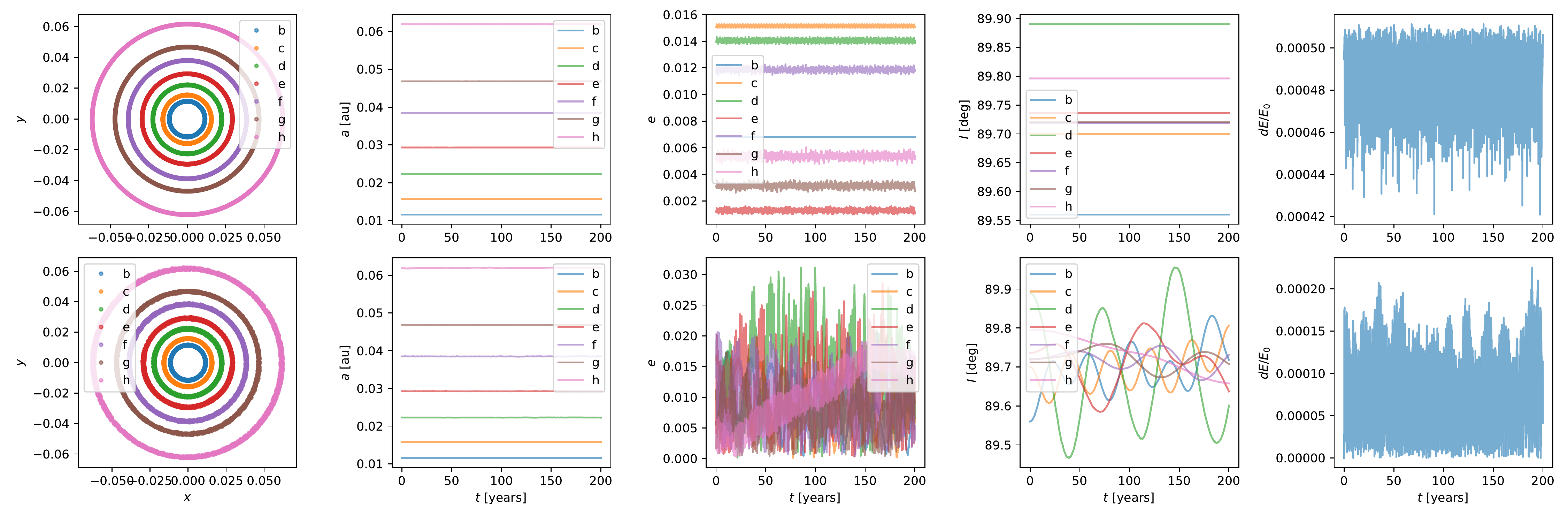}
    \caption{Evolution of the \textsc{Trappist-1} system for 200 years (equivalent to $2 \times 10^5$ time steps). First row: results from a Wisdom-Holman $N$-body integrator with neural interacting Hamiltonian; second row: ground truth results from a traditional $N$-body integrator. First column: orbit plots of Jupiter and Saturn; second column: orbital semi-major axes ($a$) as a function of time $t$; third column: orbital eccentricities $e$ as a function of $t$; fourth column: orbital inclinations $I$ (with respect to the heliocentric plane) as a function of $t$; last column: relative energy error (defined as the energy drift normalized to the initial energy) as a function of $t$. The legends ``b'', ``c'', ..., ``h'' represent the first, second, ..., and the seventh planet of the \textsc{Trappist-1} system, respectively.}
    \label{fig:trappist1}
\end{figure}

As shown in Fig.~\ref{fig:trappist1}, we integrate the system for 200 years, which corresponds to $2 \times 10^5$ integration time steps or $4.9 \times 10^4$ orbits of the innermost planet. Interestingly, even though the \textsc{Trappist-1} is well beyond the regime of the initial training data set for \texttt{WH-NIH}, the neural symplectic $N$-body integrator handles it nicely. The  evolution of the Cartesian coordinates and semi-major axis are indistinguishable from the results of the numerical integration. Concerning the $t-e$ and $t-I$ plots, the \texttt{WH-NIH} integrator  conserves the total angular momenta but fails to capture the rich dynamics of secular resonance between multiple planets. Nevertheless, we demonstrate that it is possible to learn the behavior of physical laws using a neural network and that the \texttt{WH-NIH} can apply the learned physical laws to entirely different systems. 

\section{Conclusions}
\label{sec:conclusions}
We propose an efficient neural $N$-body solver based on the splitting the symplectic Hamiltonian map, as proposed by \cite{1991AJ....102.1528W}: instead of approximating the full Hamiltonian $\mathcal{H}$, we approximate only a part of it and solve the other part analytically. In doing so, the error made by the DNN is constrained. We also develop a new activation function \texttt{SymmetricLog} that allows the neural integrator to work directly on the raw data without the need of pre-processing. We develop a custom loss function that takes into account the conservation of energy and angular momentum. We have demonstrated that our methods allows a DNN to solve the $N$-body problem for $\sim 10^5$ steps without diverging from the ground truth physics.

The proposed neural $N$-body solver is experimental, and there are some important limitations. If an $N$-body system has different interacting Hamiltonian, then the neural integrator will have to be retained. In addition, the neural integrator has been unable to understand the subtlety of secular angular momentum exchanges, and therefore it is not yet capable of modeling scenarios such as the eccentric Lidov-Kozai mechanism \citep[see, e.g.,][]{1962AJ.....67..591K,2016ARA&A..54..441N}.

\begin{ack}
It is a pleasure to thank Sam Greydanus, Javier Roa, Adrian Hamers, Vikram A. Saletore, Capsar van Leeuwen, Valeriu Codareanu, and Veronica Saz Ulibarrena for insightful discussions. This research is partially sponsored by SURF Open Innovation Lab (SOIL) and Intel Corporation.   
\end{ack}

{\small
\bibliographystyle{abbrvnat}
\bibliography{refs}
}


\appendix

\section{Appendix}
\subsection{Initial Conditions of the Simulations}
\label{appendix:ic}
To generate training data for the NIH, we carry out 50 three-body simulations with random initial conditions. For simplicity, we define the central body as a $1 M_{\odot}$\footnote{1 $M_{\odot} \approx 1.9985 \times 10^{30}$ kg is the mass of the Sun in our Solar System.} star. The masses of the two orbiters vary from 1 $M_{\odot}$ (two equal-mass stars orbit around each other, also known as a binary stellar system), $10^{-3} M_{\odot}$ (a Jupiter-mass planet orbiting around the Sun),  $10^{-6} M_{\odot}$ (an Earth-mass planet orbiting around the Sun) to $10^{-8} M_{\odot}$ (a moon-mass object orbiting around the Sun). The orbital semi-major axes ranges from 0.4 au\footnote{1 au $\approx 1.496 \times 10^8$ km is the time-averaged distance from the Sun to the Earth.} (comparable to Mercury's orbit) to 100 au (comparable to 3 times the size of Neptune's orbit). For convenience, we set $G = 4 \pi^2$ as the value of the universal gravitational constant,\footnote{When $G=4 \pi^2$ the corresponding mass unit is $M_{\odot}$, length unit is au, and time unit is years. } unless otherwise noted.

\end{document}